\documentclass[prd,onecolumn,showpacs,superscriptaddress, nofootinbib,preprintnumbers]{revtex4}
\usepackage{amsmath}
\usepackage{amsfonts}
\usepackage{graphicx}
\usepackage{hyperref}
\usepackage{mathbbol}
\def\be{\begin{equation}}
\def\ee{\end{equation}}
\def\bea{\begin{eqnarray}}
\def\eea{\end{eqnarray}}

\begin{document}
\title{Power-law expansion cosmology in Schr\"{o}dinger-type
formulation}
\date{\today}

\vspace{2cm}

\author{Burin Gumjudpai}
\email{buring@nu.ac.th}  \affiliation{Fundamental Physics \&
Cosmology Research Unit\\ The Tah Poe Academia Institute (TPTP),
Department of Physics\\ Naresuan University, Phitsanulok 65000,
Siam}

\begin{abstract}
 We investigate non-linear Schr\"{o}dinger-type formulation of cosmology
 of which our cosmological system is a general relativistic FRLW universe containing canonical
scalar field under arbitrary potential and a barotropic fluid with
arbitrary spatial curvatures. We extend the formulation to include
phantom field case and we have found that Schr\"{o}dinger wave
function in this formulation is generally non-normalizable.
Assuming power-law expansion, $a \sim t^q$, we obtain scalar field
potential as function of time. The corresponding quantities in
Schr\"{o}dinger-type formulation such as Schr\"{o}dinger total
energy, Schr\"{o}dinger potential and wave function are also
presented.
\end{abstract}

 \vspace{2pc}


\maketitle

\section{Introduction}
\label{sec:introduction} Canonical scalar field plays important
role in inflationary phase in the early universe as well as
acceleration in the late universe observed and confirmed by cosmic
microwave background \cite{Masi:2002hp}, large scale structure
surveys \cite{Scranton:2003in} and supernovae type Ia
\cite{Riess:1998cb}. The scalar field is considered as inflaton
field in inflationary models \cite{inflation}. It could also be
considered as dark energy that drives the late acceleration
described in review literatures \cite{Padmanabhan:2004av} and
references therein. In standard cosmology with
Friedmann-Lema\^{i}tre-Robertson-Walker (FLRW) background, major
components of the late universe are mixture of dark matter which
is a type of barotropic fliud and dark energy in form of scalar
field. When assuming pure scalar fluid in flat universe, one can
obtain analytical solutions otherwise the problem can also be
solved numerically. However, considering arbitrary types of
barotropic fluid and a non-flat universe, it is not always
possible to solve the system analytically.

Apart from standard cosmological equations, there are few
alternative mathematical formulations which are also equivalent to
the scalar field cosmology with barotropic fluid. One is in form of
non-linear Ermakov-attemping equation \cite{Hawkins:2001zx} and
another idea proposed recently is in form of
non-Ermakov-Milne-Pinney (non-EMP) equation. Cosmological equations
in the latter proposal can be written in form of a non-linear
Schr\"{o}dinger-type equation when imposing relations between
quantities in standard cosmological equations and
Schr\"{o}dinger-type equation \cite{D'Ambroise:2006kg}. In case of
Bianchi I scalar field cosmology, recent work shows that it is
possible to construct a corresponding linear Schr\"{o}dinger-type
equation by redefining cosmological quantities
\cite{D'Ambroise:2007gm}. With the new representation, scalar field
cosmology is reinterpreted in new way which might be able to give
new methods of approaching mathematical problems in scalar field
cosmology.

There are various observations allowing scalar field equation of
state coefficient, $w_{\phi}$ to be less than -1
\cite{Melchiorri:2002ux}. Recent data such as a combined WMAP, LSS
and SN type Ia without assuming flat universe, puts a strong
constraint, $ w_{\phi} = -1.06^{+0.13}_{-0.08}$
\cite{Spergel:2006hy}. Also the first result from ESSENCE
Supernova Survey Ia combined with SuperNova Legacy Survey Ia
assuming flat universe, gives a constraint of $w_{\phi}=-1.07\pm
0.09$ \cite{Wood-Vasey:2007jb}. Therefore it is possible that the
scalar field dark energy could be phantom, i.e $w_{\phi}<-1$
\cite{Caldwell:1999ew}. The phantom behavior, $w_{\phi}<-1$ can be
attained by negative kinetic energy term of the scalar field
density and pressure. In FLRW standard cosmology, the field can
yield big rip singularity, i.e. $a, \rho, |p| \rightarrow \infty$
at finite time \cite{Caldwell:2003vq} with attempts of singularity
avoidance in several ways \cite{Sami:2005zc}.

In this work, we investigate connection between standard
cosmological equations and non-linear Schr\"{o}dinger-type
equation with a comment on normalization of the wave function. We
modify the work of \cite{D'Ambroise:2006kg} to include phantom
field case. A case of power-law expansion with scalar field and
dark matter is considered as a toy model. We begin from Sec.
\ref{sec:cosmo} where we introduce our cosmological system.
Afterward in Sec. \ref{sec:NS}, we discuss how non-linear
Schr\"{o}dinger-type formulation quantities are related to
quantities in standard scalar field cosmology. In non-linear
Schr\"{o}dinger-type equation, one important quantity is wave
function. We comment on normalization properties of the wave
function in Sec. \ref{sec:normal}. We consider a case of power-law
expansion in Sec. \ref{sec:powerlaw} before deriving scalar field
potential, Schr\"{o}dinger potential and Schr\"{o}dinger wave
function. At last we conclude this work in Sec.
\ref{sec:conclude}.
\section{Cosmological equations} \label{sec:cosmo}
In a Friedmann-Lema\^{i}tre-Robertson-Walker universe, the
Einstein field equations are
\bea H^2 &=& \frac{\kappa^2 \rho_{\rm t} }{3} - \frac{k}{a^2}\,,
\label{fr}
\\ \frac{\ddot{a}}{a} &=& -\frac{\kappa^2}{6} (\rho_{\rm t} +
3p_{\rm t})\,, \label{ac}\eea 
where $\kappa^2 \equiv 8\pi G = 1/M_{\rm P}^2$, $G$ is Newton's
gravitational constant, $M_{\rm P}$ is reduced Planck mass, $k$ is
spatial curvature, $\rho_{\rm t}$ and $p_{\rm t}$ are total
density and total pressure, i.e. $\rho_{\rm t} = \rho_{\gamma} +
\rho_{\phi}$ and $p_{\rm t} = p_{\gamma} + p_{\phi}$. The
barotropic component is denoted by $\gamma$, while for scalar
field, by $\phi$. Equations of state for barotropic fluid and
scalar field are $p_{\gamma} = w_{\gamma}\rho_{\gamma}$ and
$p_{\phi} = w_{\phi}\rho_{\phi}$. We consider minimally couple
scalar field with Lagrangian density,
\be \mathcal{L} = \frac{1}{2} \epsilon \dot{\phi}^2 - V(\phi)\,,
\ee where $\epsilon=1$ for non-phantom case and $-1$ for phantom
case. Density and pressure of the field are given as
\bea \rho_{\phi} &=& \frac{1}{2} \epsilon \dot{\phi}^2 +
V(\phi)\,, \label{phanrho}
\\ p_{\phi}&=& \frac{1}{2} \epsilon \dot{\phi}^2 - V(\phi)\,, \label{phanp}\eea
therefore \be w_{\phi} = \frac{\epsilon \dot{\phi}^2
-2V(\phi)}{\epsilon \dot{\phi}^2 + 2V(\phi)}\,.\ee The field obeys
conservation equation
\be \epsilon\left[\ddot{\phi} + 3H\dot{\phi} \right] + \frac{{\rm
d}V}{{\rm d}\phi} = 0\,. \label{phanflu} \ee
  For the barotropic fluid, we set $w_{\gamma} \equiv
(n-3)/3$ so that $n = 3(1+w_{\gamma})$. Hence for cosmological
constant $n=0$, for fluid at acceleration bound
($w_{\gamma}=-1/3$) $n=2$, for dust $n=3$,  for radiation $n=4$,
and for stiff fluid $n=6$.  Solution of conservation equation  for
a barotropic fluid can be obtained directly by solving the
conservation equation. The solution is
\be \rho_{\gamma} = \frac{D}{a^{3(1+w_{\gamma})}} =
\frac{D}{a^{n}}\,,\label{barorho} \ee then
\be p_{\gamma} = w_{\gamma} \frac{D}{a^{n}} =
\frac{(n-3)}{3}\frac{D}{a^{n}}\,, \label{barop} \ee where a
proportional constant $D\geq 0$. Using Eqs. (\ref{fr}),
(\ref{phanrho}), (\ref{phanp}), (\ref{phanflu}) and
(\ref{barorho}), it is straightforward to show that
\bea   \epsilon \dot{\phi}(t)^2 & = & -\frac{2}{ \kappa^2} \left[ \dot{H} - \frac{k}{a^2}  \right] - \frac{n D}{3  a^n} \,, \label{phigr} \\
V(\phi) &=& \frac{3}{\kappa^2} \left[H^2 + \frac{\dot{H}}{3} +
\frac{2k}{3 a^2} \right] + \left(\frac{n-6}{6}\right)
\frac{D}{a^n}\,. \label{Vgr} \eea Therefore if one knows how the
scale factor evolves with time, the scalar field velocity and
potential can always be expressed as a function of time
explicitly.

\section{Non-linear Schr\"{o}dinger-type equation} \label{sec:NS}
Non-linear Schr\"{o}dinger-type equation corresponding to
canonical scalar field cosmology with barotropic fluid is given by
\cite{D'Ambroise:2006kg}
\bea \frac{{\rm d}^2 }{{\rm d}x^2}u(x) + \left[E-P(x)\right] u(x)
 = -\frac{nk}{2}u(x)^{(4-n)/n}\,.   \label{schroeq} \eea
 Quantities in the Schr\"{o}dinger-type equation above, e.g. wave function $u(x)$, total energy $E$ and Schr\"{o}dinger potential $P(x)$
 are related to the standard cosmology quantities as
 \bea
 u(x) &\equiv& a(t)^{-n/2}\,, \label{utoa} \\
 E &\equiv&  -\frac{\kappa^2 n^2}{12} D \,, \label{E} \\
P(x) &\equiv& \frac{\kappa^2 n}{4}a(t)^{n} \epsilon
\dot{\phi}(t)^2 \,. \label{schropotential} \eea
The mapping from cosmic time $t$ to the variable $x$ is via \be x
= \sigma(t) \label{xt}, \ee
such that \bea \dot{\sigma}(t)&=& u(x)\,, \label{dsigtou} \\
\phi(t) &=& \psi(x)\,.
 \eea

We notice that relation \be \psi'(x)^2 = \frac{4}{ \kappa^2 n}
P(x) \label{psidash} \ee in Ref. \cite{D'Ambroise:2006kg} which
gives $ \psi(x) = \pm({2}/{\kappa\sqrt{n}})
\int{\sqrt{P(x)}}\,{\rm d}x\, $ does not include phantom field
case. In order to include the phantom field case, we modify
relation
$\dot{\phi}(t) = \dot{x}\, \psi'(x) $ in \cite{D'Ambroise:2006kg}
to $ \epsilon \dot{\phi}(t)^2 = \dot{x}^2\, \epsilon\, \psi'(x)^2
 $ of which the field kinetic term ($\dot{\phi}^2$) is considered instead of the field velocity ($\dot{\phi}$) so that the parameter $\epsilon$ can be included.
   Therefore, to include the phantom field case, corrected relation to Eq. (\ref{psidash}) is \be \epsilon\,\psi'(x)^2 = \frac{4}{ \kappa^2 n}\,P(x)\,, \ee and $\psi(x)$
 should read
\bea \psi(x) =
\pm\frac{2}{\kappa\sqrt{n}}\int{\sqrt{\frac{P(x)}{\epsilon}}}\,{\rm
d}x\,  \,.
\label{phitoPx} %
\eea
 Inverse function of
$\psi(x)$ exists if $P(x) \neq 0$ and $n \neq 0$. It is important
for $\psi^{-1}(x)$ to exist as a function since existence of the
relation $x=\sigma(t)$ (Eq. (\ref{xt})) needs a
condition, %
\bea x = \psi^{-1}\circ \phi(t) = \sigma(t)\, \label{linkxt}.
 \eea
In case that $P(x) = 0$ and $n \neq 0$, then $\psi = C$, hence
inverse of $\psi$ is not a function since one $x$ gives infinite
values of $\psi^{-1}$. In this case the relation (\ref{linkxt}) is
invalid. If the inverse function, $\psi^{-1}$ exists (i.e. $P(x)
\neq 0$ and $n \neq 0$), then the scalar field potential, $V\circ
\sigma^{-1}(x)$ can be expressed as a function
of time, %
\be V(t) = \frac{12}{\kappa^2 n^2}\left( \frac{{\rm d}
u}{{\rm d}x} \right)^2 - \frac{2 u^2}{\kappa^2 n} P(x) + \frac{12
u^2}{\kappa^2 n^2}E + \frac{3 k u^{4/n}}{\kappa^2} \,.  \label{vt}
\ee
Although the potential obtained is not expressed as function of
$\phi$, however if one can integrate Eq. (\ref{phigr}) to obtain
$\phi(t)$, the obtained solution can be inserted into a known
function $V(\phi)$ motivated from fundamental physics. Then one
can check which fundamental theories give a matched potential to
$V(t)$. The Eqs. (\ref{Vgr}) and (\ref{vt}) are indeed equivalent.
Both require only the knowledge of $a(t)$, $D$ and $k$ which can
be constrained by observation. Therefore $V(t)$ in both Eqs.
(\ref{Vgr}) and (\ref{vt}) can be constructed if knowing these
observed parameters. To construct $V(t)$ in Eq. (\ref{vt}), one
needs to know $a(t)$ as a function of time in order to find $u(x)$
and $P(x)$. However, in constructing $V(t)$ in Eq. (\ref{Vgr}), if
knowing $a(t), D$ and $k$, one can directly use these quantities
without employing Schr\"{o}dinger-type quantities.

\section{Normalization condition of wave function} \label{sec:normal}
Normalization condition for a wave function $u(x)$ in quantum mechanics  is %
\be \int_{-\infty}^{\infty}|u(x)|^2 {\rm d}x = 1\,.
 \ee
The wave function here expressed as $u(x)\equiv a^{-n/2} =
\dot{x}(t)$, when applying to the normalization condition, reads%
\be \int_{-\infty}^{\infty} \dot{x}^2 {\rm d}x = 1\,.
 \ee
In order to satisfy the condition, $x$ must be constant and so is
$t$. Since the form of the wave function must be $u(x) =
\dot{x}(t)$ in order to connect equations of cosmology to the
Schr\"{o}dinger-type formulation, therefore $u(x)$ as defined is,
in general, non-normalizable.

\section{Power-law expansion} \label{sec:powerlaw}
Here in this section, we apply the method above to the power-law
expansion in scalar field cosmology with barotropic fluid in a
non-flat universe. The power-law expansion of the universe during
inflation era, \be a(t) = t^q\,, \label{powerlawex}\ee
 with $q > 1 $ was proposed by  Lucchin and Matarrese
 \cite{Lucchin} to give exponential potential
 \be
V(\phi) = \left[\frac{q(3q-1)}{\kappa^2 t_0^2 }\right]
 \exp\left\{ -\kappa\sqrt{\frac{2}{q}}
 \left[\phi(t)-\phi(t_0)\right]\right\}\,,\label{Vexp}
 \ee
assuming domination of scalar field, negligible radiation density
and negligible spatial curvature. Recent results from X-Ray gas of
galaxy clusters put a constraint of $q \sim 2.3$ for $k = 0$, $q
\sim 1.14$ for $k = -1$ and $q \sim 0.95$ for $k = 1$
\cite{Zhu:2007tm}. Considering mixture of both fluids, we use
effective equation of state,
$ w_{\rm eff} = ({\rho_{\phi}w_{\phi} + \rho_{\gamma}
w_{\gamma}})/{\rho_{\rm t}}.$ For a flat universe, the power law
expansion, $a= t^q$, is attained when $-1< w_{\rm eff} < -1/3$
where $q = 2/[3(1+w_{\rm eff})]$. If using $q=2.3$ as mentioned
above, it gives $w_{\rm eff} = -0.71$.

\subsection{Relating Schr\"{o}dinger quantities to standard
cosmological quantities}%
 Assuming power-law expansion and using
Eqs. (\ref{utoa}) and (\ref{dsigtou}),
Schr\"{o}dinger wave function is related to standard cosmological quantity as %
\be
u(x) = \dot{\sigma}(t) = t^{-qn/2}   \,. \label{u(x)} \ee We can
integrate the equation above so that the Schr\"{o}dinger scale,
$x$
is related to cosmic time scale, $t$ as %
\be x = \sigma(t) = -\frac{t^{-\beta }}{\beta} + \tau\,,
\label{xtot}\ee where $\beta \equiv (qn-2)/2 $ and $\tau$ is an
integrating constant. The parameters $x$ and $t$ have the same
dimension since $\beta$ is only a number. Using Eq.
(\ref{powerlawex}), we can find $\epsilon\dot{\phi}(t)^2$ from Eq.
(\ref{phigr}):
\be \epsilon\dot{\phi}(t)^2 = \frac{2q}{\kappa^2 t^2}  +
\frac{2k}{\kappa^2 t^{2q}} - \frac{n D}{3 t^{qn}}\,.
\label{phigrpowerex}\ee
We use
 Eqs.  (\ref{powerlawex}) and (\ref{phigrpowerex}) in Eq. (\ref{schropotential}), therefore
the Schr\"{o}dinger potential is found to be %
\be%
P(x) = \frac{qn}{2}t^{qn-2} + \frac{k n}{2}t^{q(n-2)} -
\frac{\kappa^2 n^2 D}{12} \,. \label{Px}
 \ee
With $E = -\kappa^2 n^2 D/12$, the Schr\"{o}dinger kinetic energy
is \be T = -\frac{qn}{2}t^{qn-2} - \frac{k n}{2}t^{q(n-2)}\,.
\label{keterm1} \ee

 \subsection{Scalar field potential $V(t)$}
\begin{figure}[t]
\begin{center}
\includegraphics[width=6.5cm,height=6.4cm,angle=0]{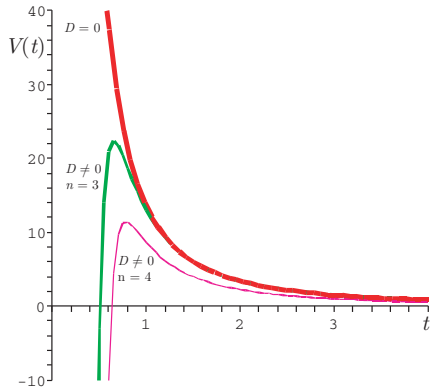}
\end{center}
\caption{Potential $V(t)$ plots from non-linear
Schr\"{o}dinger-type formulation assuming $a\sim t^q$, $q=2.3$ in
flat universe ($k=0$). The thickest line is when there is no
barotropic fluid $D=0$. The middle line is when there is dust
fluid together with scalar field, i.e. $D \neq 0$ and $n=3$. The
small line is when the universe has scalar field with radiation
fluid, i.e. $D \neq 0$ and $n=4$. We set $\kappa = 1$ and in the
last two plots, we set $D= 1$. All plots match results obtained
from standard cosmological equations. \label{pic1}}
\end{figure}

\begin{figure}[t]
\begin{center}
\includegraphics[width=5.4cm,height=4.7cm,angle=0]{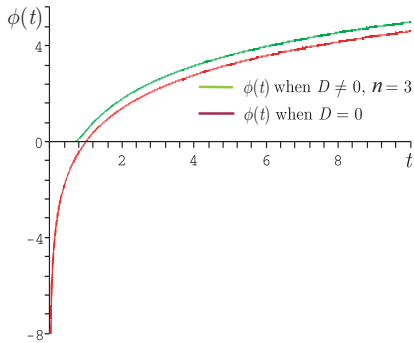}
\end{center}
\caption{$\phi(t)$  for power-law expansion $a\sim t^q$, $q=2$ in
flat universe ($k=0$). The red line is of the when the barotropic
fluid density is negligible. The green line is in the presence of
scalar field with dust ($D \neq 0$ and $n=3$). In the figure,
$\kappa = 1$ and $D= 1$. \label{pic2}}\end{figure}

\begin{figure}[t]
\begin{center}
\includegraphics[width=5.9cm,height=4.9cm,angle=0]{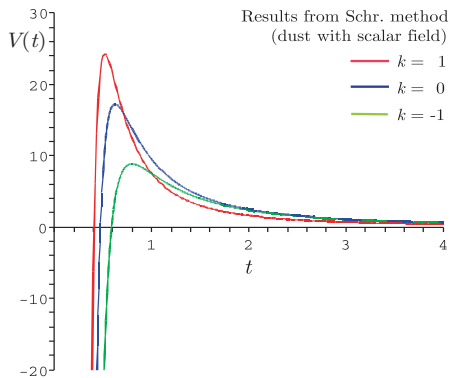}
\end{center}
\caption{ $V(t)$ obtained from non-linear Schr\"{o}dinger-type
formulation for closed, flat and open universe in presence of dust
and scalar field. \label{pic3}}
\end{figure}

In order to obtain $V(t)$ in Eq. (\ref{vt}), we need to
know derivative of $u(x)$: %
\bea \frac{{\rm d}}{{\rm d}x}\,u(x) &=&   - \frac{qn}{2t}\,.
\label{udash}  \eea
 At this step,
using Eqs. (\ref{utoa}), (\ref{E}), (\ref{schropotential}) and
(\ref{udash}) in Eq. (\ref{vt}), we finally obtain
\be V(t) = \frac{q(3q-1)}{\kappa^2 t^2} + \frac{2k}{\kappa^2
t^{2q}} + \left(\frac{n-6}{6}\right)\frac{D}{t^{qn}}\,.
\label{vtpowerlaw}\ee
Assuming flat universe ($k=0$) and $q=2.3$, we show $V(t)$ in Fig. \ref{pic1}. Thickest line on top is of the case scalar field without
barotropic fluid. The middle line is the case when the dust is presented with scalar field ($D \neq 0, n = 3$). The bottom line is the case of
radiation ($D \neq 0, n = 4$). The $V(t)$ plots from the Schr\"{o}dinger-type formulation matches the plots from standard cosmological
equations. The result is independent of $\epsilon$ values.  The solution $\phi(t)$ of Eq. (\ref{phigrpowerex}) can not be integrated if
$\epsilon = -1$ or if the integrand of Eq. (\ref{phigrpowerex}) is imaginary. When $\epsilon = 1$ with dust ($D \neq 0, n=3$) and $q=2.3$, the
integrand is imaginary. We therefore assume $q=2$ to show numerical integrations in Fig. \ref{pic2} for the case $D=0, k=0$ and the case $D\neq
0, n=3, k=0$. In the pure scalar field case $D=0, k=0$, numerical solution matches the analytical solution $\phi(t) = (\sqrt{2q}/\kappa)\ln(t)$.
This solution can be substituted into Eq. (\ref{vtpowerlaw}) to obtain Eq. (\ref{Vexp}) as in \cite{Lucchin} (setting $t_0=1$ and $\phi(t_0) =
0$). When considering cases of closed, flat and open universe containing dust matter, $V(t)$ of each case is presented in Fig. \ref{pic3} where
$q=2$ is assumed in all cases so that we can see how the plots change their shapes when $k$ is varied. It is worth noting that reconstruction of
scalar field potential assuming scaling solution was considered before in \cite{Rubano:2001}.

\subsection{Schr\"{o}dinger potential $P(x)$}
\begin{figure}[t]
\begin{center}
\includegraphics[width=7.2cm,height=9.7cm,angle=0]{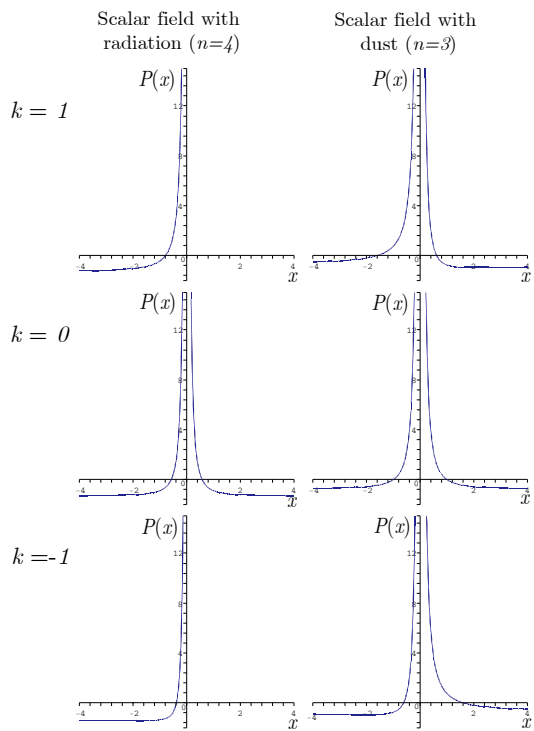}
\end{center}
\caption{$P(x)$ plotted versus $x$ for power-law expansion. We set
$q=2, \kappa=1, D=1 $ and $ \tau=0$. The scalar field dominant
case can not be plotted because even tough we set a condition
$D=0$, the coefficient $n$ of the barotropic fluid equation of
state still appears in the first and second terms of the Eq.
(\ref{Pxpower}). There is only a real-value $P(x)$ for the cases
$k=\pm 1$ with $n=4$ because, when $x>0$, $P(x)$ becomes imaginary
in these cases.} \label{pic4}
\end{figure}

We can find Schr\"{o}dinger potential $P(x)$ from Eqs.
(\ref{xtot}) and (\ref{Px}) where time is expressed as a function
of $x$ as 
\be t(x) = \frac{1}{\left[-\beta (x-\tau) \right]^{1/\beta}}
\label{ttox}\,. \ee Therefore
\bea
P(x) &=& \frac{2qn}{(qn-2)^2}\frac{1}{(x-\tau)^2}  +\:
\frac{kn}{2}\left[ \frac{-2}{(qn-2)(x-\tau)}
\right]^{2q(n-2)/(qn-2)}    - \: \frac{\kappa^2 n^2 D }{12}\,.
\label{Pxpower}
\eea As in Eq. (\ref{keterm1}), the Schr\"{o}dinger kinetic energy
is
\bea T(x) &=&  - \: \frac{2qn}{(qn-2)^2}\frac{1}{(x-\tau)^2} -\:
\frac{kn}{2}\left[ \frac{-2}{(qn-2)(x-\tau)}
\right]^{2q(n-2)/(qn-2)}. \eea %
The kinetic term has contribution only from the power $q$ and
spatial curvature $k$. A disadvantage of Eq. (\ref{Pxpower}) is
that we can not use it in the case of scalar field domination as
in inflationary era. Dropping $D$ term in Eq. (\ref{Pxpower}) can
not be considered as scalar field domination case since the
barotropic fluid coefficient $n$ still appears in the other terms.
The non-linear Schr\"{o}dinger-type formulation is therefore
suitable when there are both scalar field and a barotropic fluid
together such as the situation when dark matter and scalar field
dark energy live together in the late universe. The
Schr\"{o}dinger potentials $P(x)$ plotted with $x$ for power-law
expansion with $q=2$ in closed, flat and open universe are shown
in Fig. \ref{pic4}. In the figure, the dust cases are shown on the
right and radiation cases are on the left. We set $\kappa=1, D=1$
and $ \tau=0$.

\subsection{Schr\"{o}dinger wave function $u(x)$}
\begin{figure}[t]
\begin{center}
\includegraphics[width=5.4cm,height=5.5cm,angle=0]{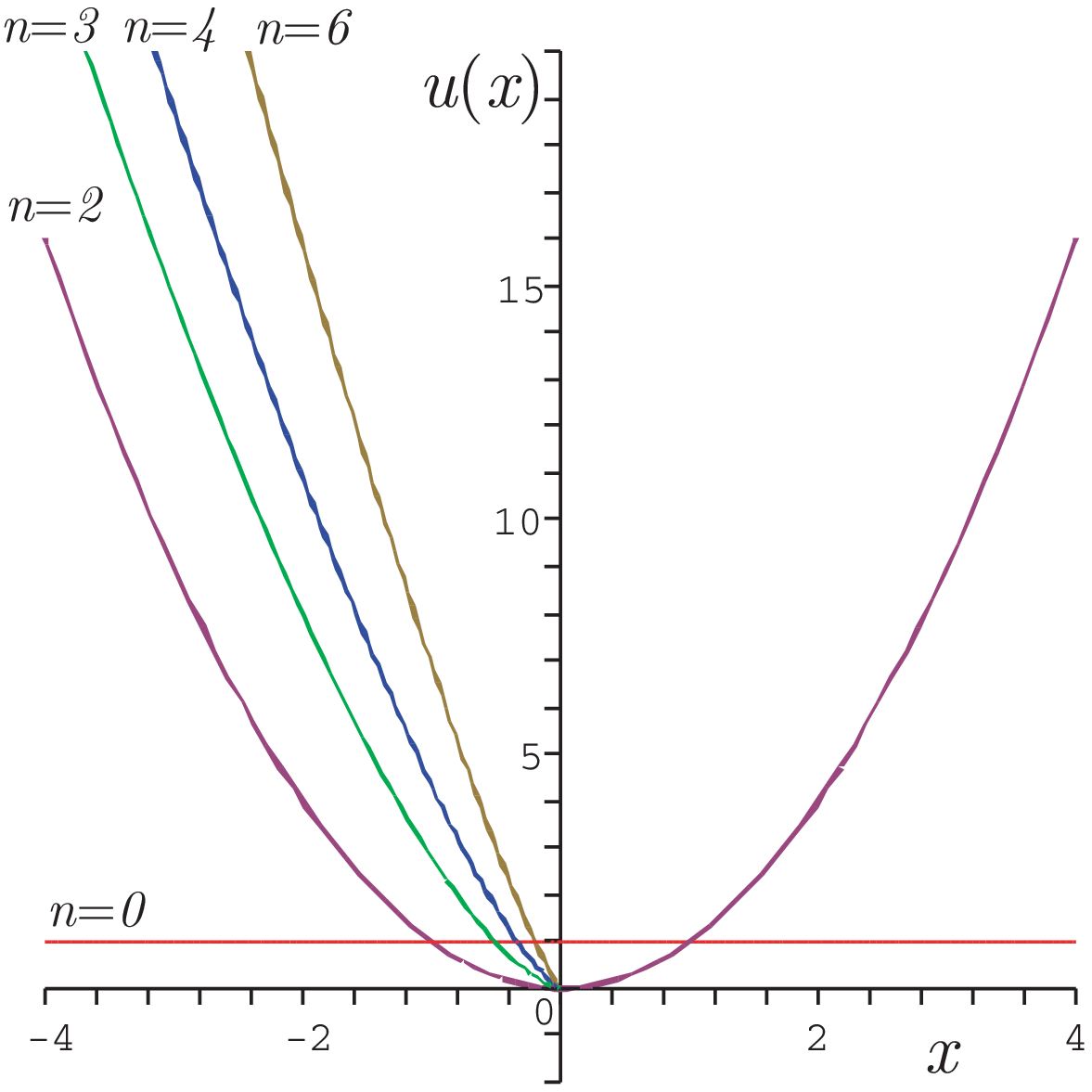}
\end{center}
\caption{$u(x)$ plotted versus $x$ for power-law expansion with
$q=2$. We set $ \tau=0$. The wave function is plotted for $n=0$
(cosmological constant), $n=2$, $n=3$ (dust), $n=4$ (radiation)
and $n=6$ (stiff fluid). There is no real-value wave function  for
$n=3$, $n=4$ and  $n=6$ unless $x<0$.} \label{pic5}
\end{figure}

The quantity analogous to Schr\"{o}dinger wave function can be
directly found from Eqs. (\ref{u(x)}) and (\ref{ttox}) as
\be u(x) = \left[  \left( -\frac{1}{2}qn+1 \right)(x-\tau)
\right]^{qn/(qn-2)} \,, \label{uxt} \ee which is independent of
the spatial curvature $k$ or the initial density $D$. However,
coefficient $n$ of the barotropic fluid equation of state and $q$
must be expressed.  Wave functions for a range of barotropic
fluids are presented in Fig. \ref{pic5}. The result is confirmed
by substituting Eq. (\ref{uxt}) into Eq. (\ref{schroeq}).


\section{Conclusions and Comments} \label{sec:conclude}

We consider Schr\"{o}dinger-type formulation for a system of
canonical scalar field and a barotropic fluid in standard FLRW
cosmology with zero or non-zero spatial curvature. In the
Schr\"{o}dinger-type formulation, all quantities in cosmology are
represented in Schr\"{o}dinger-like quantities and the equation
relating these Schr\"{o}dinger-like quantities is written as a
non-linear Schr\"{o}dinger-type equation.   If $a(t)$ is known as
an exact function of time, a connection of two scale quantities,
$x$ and $t$ can be found and then other Schr\"{o}dinger-like
quantities can be determined. We modified the formulation to
include the phantom field case. The equation can be simplified to
linear type if we consider the flat universe case $k=0$ or the
cases $n=2$ or $n=4$ \cite{D'Ambroise:2006kg}. However, even if
the equation is linear, it can not be considered as an analog to
non-relativistic time-independent quantum mechanics because in
this work, the wave function of Schr\"{o}dinger-type formulation
is found to be, in general, non-normalizable. Afterward, we
consider a particular case of power-law expansion of scale factor.
We show relations between cosmological quantities in conventional
form and in Schr\"{o}dinger-like form for power-law expansion. We
obtain scalar field potential $V(t)$, Schr\"{o}dinger potential
$P(x)$ and wave function $u(x)$. In the case of a scalar field
dominant in flat universe, our analytical results $V(\phi)$ and
$\dot{\phi}$ agree well with the well-known results in
\cite{Lucchin}.  A range of plots in various cases of closed, flat
or open geometries is presented. Wave functions for the power-law
expansion case (seen in the Fig. \ref{pic5}) are found to be all
non-normalizable as conjectured.

Without knowledge of $a(t)$, one might wonder if we could start
the calculation procedure from solving the Schr\"{o}dinger-type
equation (\ref{schroeq}) for example, the linear case as done in
basic quantum mechanics. However, in order to do this, we must
know the Schr\"{o}dinger potential $P(x)$ (Eq.
(\ref{schropotential})) which depends explicitly on $a(t)$ and
$\dot{\phi}$. Nevertheless, $\dot{\phi}$ (Eq. (\ref{phigr})) also
depends on $a(t)$. Therefore we need to know the law of expansion
$a(t)$ before proceeding the calculation. Knowing $a(t)$ enables
us to know $u(x)$ directly (see Eq. (\ref{u(x)})). Hence in
Schr\"{o}dinger-type formulation, we do not work as in basic
quantum mechanics in which major task is to solve the
Schr\"{o}dinger equation for $u(x)$. There could be many solutions
of a Schr\"{o}dinger-type equation. In quantum mechanics valid
solutions $u(x)$ must be only normalizable type. Here, unlike in
quantum mechanics, our $u(x)$ must be non-normalizable.

At late time the scalar field dark energy and cold dark matter
(dust) are two major components of the universe while the others
are negligible. For power-law expansion, the procedure is suitable
for studying the system of scalar field dark energy and dark
matter because it gives all real-value of $P(x)$ for any $k$. We
need to know $a(t)$, $k$ and $D$ which are observable in order to
find $V(t)$. Information of $V(\phi)$ is important because it is a
link to fundamental physics. If one starts from fundamental
physics with a particular potential $V(\phi)$ and if also knowing
how $\phi$ evolves with $t$, then $V$ could be expressed as
function of $t$. Finally, the potential $V(t)$ obtained from
observation and another $V(t)$ proposed by fundamental physics can
be compared. The non-linear Schr\"{o}dinger-type formulation might
provide an alternative mathematical approach to problem solving in
scalar field cosmology.

\section*{Acknowledgements}
 B. G. is a TRF Research Scholar under a TRF-CHE Research
Career Development Grant of the Thailand Research Fund and the
Commission on Higher Education of Thailand. This work is also
supported by Naresuan Faculty of Science Research Scheme.

\end{document}